\documentclass[superscriptaddress,nofootinbib,twocolumn]{revtex4}
\usepackage{amsmath,lscape,epsfig}
\usepackage{amsfonts}
\usepackage{amsmath}
\usepackage{amssymb}
\usepackage{graphicx}

\def\ii{\'{\i}}
\def\beq{\begin{equation}}
\def\eeq{\end{equation}}
\def\beqa{\begin{eqnarray}}
\def\eeqa{\end{eqnarray}}
\def\ban{\begin{eqnarray*}}
\def\ean{\end{eqnarray*}}
\def\bi{\begin{itemize}}
\def\ei{\end{itemize}}

\begin{document}

\title{Quark matter under strong magnetic fields in the Nambu--Jona-Lasinio 
Model}

\author{D.P. Menezes}
\affiliation{Depto de F\'{\i}sica - CFM - Universidade Federal de Santa
Catarina  Florian\'opolis - SC - CP. 476 - CEP 88.040 - 900 - Brazil}
\author{M. Benghi Pinto}
\affiliation{Depto de F\'{\i}sica - CFM - Universidade Federal de Santa
Catarina  Florian\'opolis - SC - CP. 476 - CEP 88.040 - 900 - Brazil}
\author{S.S. Avancini}
\affiliation{Depto de F\'{\i}sica - CFM - Universidade Federal de Santa
Catarina  Florian\'opolis - SC - CP. 476 - CEP 88.040 - 900 - Brazil}
\author{A. P\'erez Mart\'inez}
\affiliation{Instituto de Cibern\'etica Matem\'atica y F\'isica (ICIMAF) -
Calle E esq 15 No. 309 Vedado, Havana, 10400, Cuba}
\author{C. Provid\^encia}
\affiliation{Centro de F\ii sica Computacional - Department of Physics -
University of Coimbra - P-3004 - 516 - Coimbra - Portugal}

\begin{abstract}

In the present work we use the large-$N_c$ approximation to investigate quark matter 
described by the SU(2)  Nambu--Jona-Lasinio model subject to a strong magnetic field. The Landau
levels are filled in such a way that usual kinks  appear in the effective
mass and other related quantities. $\beta$-equilibrium is also considered and
the macroscopic properties of a magnetar described by this quark matter
is obtained. Our study shows that the magnetar masses and radii are 
larger if the magnetic field increases but only very large fields 
($\ge 10^{18}$ G) affect
 the EoS in a non negligible way.
\end{abstract}

\maketitle

\vspace{0.50cm}
PACS number(s): {24.10.Jv,26.60+c,11.10.-z,11.30Qc}
\vspace{0.50cm}

\section{Introduction}

In 1979, telescopes in spacecrafts and astronomers around the world detected
the emission of very intense gamma and X rays. The sources of these rays
were first called soft gamma repeaters (SGR) and later identified as possible
remnants of supernova explosions, the tragic death of very massive stars.
If this remnant is a neutron star that spins very rapidly, an intense
magnetic field is formed; if it spins slower the magnetic field is
strong but not as much as in the first case. The ordinary neutron stars, also
known as pulsars, bear a magnetic field of the order of $10^{12}-10^{13}$ G.
The neutron stars with very strong magnetic fields of the order of
$10^{14}-10^{15}$ G are known as magnetars and they are believed to be the
sources of the intense gamma and X rays detected in 1979. Most of the time the
magnetar remains
inactive, but the strong magnetic field causes the solid crust to break into
small pieces. The crustquake  leaves a fireball which
cools down and emits X rays from
its surface until it evaporates completely \cite{duncan,kouve}.

In Ref. \cite{prakash} the equation of state used to describe neutron stars in
a strong magnetic field is obtained from a field theoretical approach. Two
relativistic models are used: one normally called non-linear Walecka model
(NLWM) and the other one with a derivative coupling between mesons and baryons.
The importance of including anomalous magnetic moments (AMM) is discussed.
A more recent work \cite{wei} analyses the importance of the scalar-isovector
$\delta$ mesons in the EoS that describes magnetars. In \cite{cp} density
dependent hadronic models are used with the same purpose. In all three works
the AMM of the electrons were not considered because they were shown to cancel
out if properly introduced \cite{duncan}.  In two papers \cite{prakash,wei}
the AMM of the muons were also taken into account. All above mentioned papers
refer to neutron stars composed of hadrons and leptons. These stars are called
hadronic stars.

In the stellar modeling, the structure of the star depends on the
assumed equation of state built with appropriate models. The true
ground state  of matter remains a source of speculation. In
conventional models, hadrons are assumed to be the true ground state
of the strong interaction. However, it  has been argued
\cite{itoh,bodmer,witten,haensel,olinto} that {\it strange quark
matter} (SQM) is the true ground  state of all matter. This hypothesis is
known as the Bodmer-Witten conjecture. Hence, the
interior of neutron stars should be composed predominantly of
$u,d,s$ quarks, plus leptons to ensure charge neutrality. Pulsars
described by matter composed of SQM are often called strange stars.
However, as the strangeness content depends on the model used to
describe the quark matter, we prefer to describe any model in which
the interior involves deconfined quarks (not bound in hyperons) as
{\it quark stars} \cite{quarks}. Apart from the differences in the
EoS, an important distinction between quark stars and conventional
neutron stars is that the quark stars are self-bound by the strong
interaction, whereas neutron stars are bound by gravity. This allows
a quark star to rotate faster than would be possible for a neutron
star. Moreover, some authors have argued that quark stars should be
bare \cite{xureview,usov}, in the sense that any  crust would either
not form or would be destroyed during the supernova explosion. The
characteristics of the radiation from hot, bare strange stars have
been identified \cite{usov04} and the electron-positron pairs that
can be emitted from bare quark stars \cite{pairs} require the
existence of a surface layer of electrons tied to the star by a
strong electric field. Two of the most common models used to
describe quark matter are the MIT bag model \cite{bag} and the
Nambu--Jona-Lasinio model (NJL) \cite{njl}. In \cite{quarks} it was
shown that while the electron chemical potential of a quark star
described by the MIT bag model is very low (less than 20 MeV), the
NJL model gives much higher values reaching 100 MeV inside the star,
accounting for the necessary electric field that explains the
emission of electron-positron pairs.

An important point to be investigated refers to the stability of quark matter
in the interior of quark stars. Two different possibilities for the MIT bag
model can be found in relation to quark matter in the interior of quark stars:
the unpaired phase \cite{bag}, which is widely favored in the literature on
strange stars, and the color-flavor-locked phase (CFL) \cite{cfl},
which allows the quarks near the Fermi surface to form Cooper pairs
which condense and break the color gauge symmetry \cite{mga}. At
sufficiently high densities the favored phase is the CFL.
In \cite{klimenko03} the stability of quark matter described by the two-flavor
NJL model and subject to an external magnetic field was investigated. It was
shown that the stability depends on the strength of this field and on model
parameters.

In Ref. \cite{aurora} the EoS for magnetized quark stars was described with the
help of the MIT bag model. AMM for the quarks were properly taken into account.

The NJL model in a magnetic field was considered in \cite{Ghosh:2005rf} and
it was shown that the magnetic field spontaneously breaks chiral symmetry. The
formalism used in the above mentioned study was based on a previous calculation
performed in Ref. \cite{providencia}.

The scope of the present work is to study $ud$ quark matter in a magnetic
  field within the NJL model, with or without the requirement of
$\beta$-equilibrium.
We focus our work on the SU(2) version of the model.
Moreover, we remark that the present work may also be relevant
regarding the physics of non-central heavy ion collisions such as
the ones performed at RHIC and LHC-CERN which can provide a possible
signature for the presence of CP-odd domains in the presumably
formed quark-gluon plasma phase \cite {qgp}. In this particular case
one reaches magnetic fields of about $10^{19} \, {\rm G}$ or $B \simeq
6m_\pi^2/e$ ($m_\pi$ representing the pion mass and $e$ the
fundamental electric charge).

This work is organized as follows: In section II we set up a
lagrangian density adequate to describe two flavor quark matter, in
$\beta$ equilibrium, in the presence of an external magnetic field.
This allows the  derivations  to be carried out in a uniform
fashion from a common lagrangian density making the evaluations more
transparent and pedagogically more interesting to the reader. Using
functional techniques we derive the effective potential, which
represents Landau's free energy density, using the standard
large-$N_c$ approximation. In section III the relevant expressions
for quark matter EoS subject to a magnetic field are displayed.
Specific cases, for the symmetric version, are discussed in section
IV A for finite density and no magnetic field, IV B for zero density
in a magnetic field and IV C for the general case of finite density
matter in a magnetic field. In section V, details of matter in
$\beta$- equilibrium are given and in section VI our results are
shown and discussed. In section VII the more important conclusions
are drawn.

\section{General formalism}

In order to consider (two flavor) quark stars in $\beta$ equilibrium with
strong magnetic fields
one may  define the following lagrangian density
\begin{equation}
{\cal L}_{\beta f} = {\cal L}_{f}+{\cal L}_{l} - \frac {1}{4}F_{\mu
\nu}F^{\mu \nu}
\end{equation}
where the quark sector is described by the  Nambu--Jona-Lasinio model
\begin{eqnarray}
{\cal L}_f &=& {\bar{\psi}}_f \left[\gamma_\mu\left(i\partial^{\mu}
- q_f A^{\mu} \right)-
m_c \right ] \psi_f \nonumber \\
&+& G \left [({\bar \psi}_f \psi_f)^2 + ({\bar \psi}_f i\gamma_5
{\vec \tau} \psi_f)^2 \right ]\;, \label{njl}
\end{eqnarray}
where a summation over the quark flavors, $f=u,d$ is implied while $q_f$ represents the quark electric charge.

Note that we have used $m_c = m_u \simeq
m_d$ as representing the current masses.

The leptonic sector is given by
\begin{equation}
\mathcal{L}_l=\bar \psi_l\left[\gamma_\mu\left(i\partial^{\mu} - q_l A^{\mu}
\right) -m_l\right]\psi_l \,\,,
\label{lage}
\end{equation}
where $l=e,\mu$. One recognizes this sector as being represented by
the usual QED type of lagrangian density. As  usual, $A_\mu$ and $F_{\mu \nu }=\partial
_{\mu }A_{\nu }-\partial _{\nu }A_{\mu }$ are used to account
for the external magnetic field. Then, since we are interested in a
  static and constant magnetic field
in the $z$ direction, $A_\mu=\delta_{\mu 2} x_1 B$.

Regarding the actual evaluations,
${\cal L}_f$ and ${\cal L}_l$ bear two fundamental differences. Due
to the quadratic fermionic interaction, the former is non
renormalizable in 3+1 dimensions ($G$ has dimensions
of ${\rm eV}^{-2}$ ), meaning that eventual divergences
cannot be eliminated by a consistent redefinition of the original model
parameters (fields, masses, and couplings). The renormalizability issue arises during the
evaluation of momentum integrals which represent Feynman loops and, in the process, one usually employs
regularization prescriptions (e.g. dimensional regularization, sharp cut-off, etc) which are  formal ways to isolate divergences. However, the procedure introduces {\it arbitrary } parameters with dimensions of energy  which do not appear
in the {\it original} lagrangian density. Later, when (unlike the 3+1 NJL model) the theory is renormalizable, one may choose
any value for the arbitrary energy scale and the original parameters  {\it run} with it as dictated by
the renormalization group.
Within the NJL model a sharp cut off ($\Lambda$) is preferred and since the model is nonrenormalizable one gives up
the very high energy scales fixing $\Lambda$ to a value related to the physical spectrum under investigation. This strategy
turns the 3+1 NJL model into an effective model while $\Lambda$ is treated as
a {\it parameter}. The experimental values
of quantities such as the pion mass ($m_{\pi}$) and the pion decay
constant $(f_{\pi})$ are used to fix both, $G$ and $\Lambda$.

 A second important issue regards the fact that, when $m_c \to 0$ , the quark
propagator brings unwanted infra red divergences meaning that the evaluations have to be
carried out in a {\it nonperturbative} fashion. Moreover, very often physical quantities
(such as the self energy)
appear as powers of the dimensionless quantity $G \Lambda^2$ which is greater than the unity
spoiling any possibility of success via standard perturbative evaluations.

As far as analytic nonperturbative evaluations are concerned, one can consider one loop contributions
dressed up by a fermionic propagator whose effective mass ($M$) is
determined in a self consistent way. This approximation is known
under different names, e.g., Hartree, large-$N_c$ or mean field
approximations (MFA). The leptonic sector, on the other hand, is described
by a QED type of lagrangian density which is renormalizable and which, in principle, could   be
treated in a perturbative fashion. Here,  we  treat the complete
${\cal L}_{\beta f}$ in a consistent way by evaluating the
thermodynamical potential related to  ${\cal L}_{\beta f}$ up to
one loop. It is interesting to remark that, in practice, one does not
have to consider the full ${\cal L}_{\beta f}$ since both sectors
have similar polynomial structures, apart from the  four quark
interaction term which, as discussed, makes the theory
nonrenormalizable and renders perturbative calculations useless. So,
concerning the evaluation of the equation of state, our strategy is
the following. We use quantum field methods, in the imaginary
time formalism, to evaluate the effective potential, or Landau's
free energy density (${\cal F}_f$), for the quark sector in the large-$N_c$ approximation.
  After summing over the Matsubara's frequencies we will regularize the divergent (three) momentum integrals
by using a sharp {\it non covariant} cut off, $\Lambda$.
When evaluated at its minimum, the effective potential is
equivalent to the thermodynamical potential,
$\Omega_f = -P_f = {\cal E}_f - T {\cal S} - \mu_f \rho_f $ where $P_f$
represents the pressure, ${\cal E}_f$ the energy density, $T$ the temperature,
${\cal S}$ the entropy density, and
$\mu_f$ the chemical potential (a sum over repeated indices is implied).  The quark density, $\rho_f$, and the
 baryonic density, $\rho_B$, are simply related by $\rho_f = 3 \rho_B$.
For the present study, just the zero temperature
case is important and, as a consequence, the term with the entropy vanishes.
Equivalent results for the
leptonic sector, within the same approximation, can be trivially obtained by performing the
replacements $G \to 0$, $m_c \to m_l$, $q_f \to q_l$ and $N_c \to 1$. Finally, this procedure allows us to obtain
the EoS from the full pressure for $\beta$ stable dense quark matter
in the presence of a magnetic field($B$)
\begin{equation}
P_{\beta f}(\mu_f,\mu_l,B)= P_f (\mu_f,B)|_{M}+ P_l(\mu_l,B)|_{m_l} +\frac{B^2}{2}
\,\,,
\end{equation}
where our notation means that $P_f$ is evaluated in terms of  the quark effective mass, $M$, which is
determined in a (nonperturbative) self consistent way while $P_l$ is evaluated at the leptonic bare mass, $m_l$. The term  $B^2/2$ arises due to the electromagnetic  term $F_{\mu \nu}F^{\mu \nu}/4$ in the
original lagrangian density. In order to normalize our results we shall require that the pressure vanishes at zero chemical potentials by defining
\begin{equation}
P_{\beta f, {\rm eff}}(\mu_f,\mu_l,B)=P_{\beta f}(\mu_f,\mu_l,B)-P_{\beta f}(0,0,B) \,\,.
\label{completeP}
\end{equation}
Note that the above normalization prescription, which washes away the $B^2/2$ term, is not unique and one 
could as well require that the pressure vanishes at zero chemical potential
and zero magnetic field in which case the $B^2/2$ term  survives. Although we
use the former prescription for most of the time we will also consider the
latter in  section VI, when stellar matter is discussed.
Throughout this paper we consider the following set of parameters
\cite {buballa}:  $\Lambda = 587.9 \, {\rm MeV}$ , $m_c= \,  5.6 {\rm MeV}$,
$m_e=0.511 \, {\rm MeV}$ , $m_\mu=105.66 \, {\rm MeV}$, and $G \Lambda^2=2.44$.

\section{EoS for quark matter at finite density in a magnetic field}

To obtain the effective potential (or Landau free energy density)
for the quarks, ${\cal F}_f$, it is convenient to consider the
bosonized version of the NJL which is easily obtained by introducing
auxiliary fields ($\sigma,{\vec \pi}$) through a
Hubbard-Stratonovich type of transformation. Here, ${\cal F}_f$ is
evaluated using the large-$N_c$ approximation which is equivalent to
the mean field approximation (MFA). Then, to introduce the auxiliary
bosonic fields and to render the theory more suitable to apply the
large-$N_c$ approximation it is convenient to use $G \to \lambda/(2
N_c)$ formally treating $N_c$ as a large number which is set to
its relevant value, $N_c=3$, at the end of the evaluations. To retrieve some well known
results related to chiral symmetry breaking (CSB), in the absence of a magnetic field,
let us set $B=0$ for the moment \footnote{ As we shall see in the sequel one may easily incorporate
contributions for $T$, $\mu$ and $B$ by modifying some of the relevant Feynman rules}.
 One then has
\begin{equation}
{\cal L}_f = {\bar \psi}_f \left( i \not\!\partial\right ) \psi_f
-{\bar \psi}_f(\sigma^\prime + i\gamma_5 {\vec \tau}{\vec
\pi})\psi_f -\frac{N_c}{2 \lambda}(\sigma^2 + {\vec \pi}^2) \;,
\label{njlboson}
\end{equation}
where $\sigma^\prime=\sigma+m_c$.
Note that the introduction of the auxiliary (or background) fields does not
change the
physics since they do not propagate and their Euler-Lagrange equations of motion trivially lead
to $\sigma = - (\lambda/ N_c) \bar \psi_f \psi_f = - 2G \bar \psi_f \psi_f$ and
${\vec \pi} = - 2G i \bar \psi_f \gamma_5 {\vec \tau} \psi_f$.
However, their introduction simplifies the selection of the relevant contributions at any order
in $1/N_c$ \cite {coleman}. Basically, each closed quark loop contributes with a factor of $N_c$ while
each internal bosonic line brings a factor of $1/N_c$.
The effective potential (or Landau's free energy density), ${\cal F}_f$, is defined as the classical potential
which appears (at the tree level) in the original plus radiative (quantum) corrections. As it is well known,
${\cal F}_f$ is particularly useful in the study of symmetry breaking/restoration since a symmetry which is
observed at the classical level may be broken by quantum corrections with the appearance of a non vanishing order
parameter. The effective potential, which is also the  generating functional all 1PI contributions
with zero external momenta \cite {coleman},  can be readily obtained by integrating over the fermionic fields.
Within the large-$N_c$ approximation this procedure results in \cite {klimenko}
\begin{equation}
{\cal F}_f= \frac {N_c (\sigma^2+\pi^2)}{2
\lambda} + i  \int \frac {d^4 p}{(2\pi)^4} {\rm tr}\ln \left[\not \!
p - ( \sigma^\prime+ i\gamma_5 {\vec \tau}{\vec \pi})\right] \,\,.
\label{general}
\end{equation}
The first term on the right hand side is the classical (tree)
contribution while the second accounts for a radiative (loop)
contribution of order-$N_c$. Let us  define some important physical
quantities by quickly reviewing how CSB
arises within the NJL model. For this let us take the chiral limit
($m_c=0$). Equation (\ref {general}) can be written as

\begin{equation}
{\cal F}_f(\chi)= \frac {N_c \chi^2}{2 \lambda} + \frac{i}{2}  {\rm tr}  \int  \frac {d^4 p}{(2\pi)^4} \ln \left[ -p^2 + \chi^2 \right]
 \,\,,
 \label{gentrace}
\end{equation}
where we have  defined $\chi= \sqrt {\sigma^2 + \pi^2}$. Since the potential is symmetric
we can set $\vec \pi=0$ considering only the $\sigma$ direction. Minimizing with respect to
$\sigma$ leads to the well known gap equation
\begin{equation}
\frac{ d{\cal F}_f}{ d\sigma} {\Big |}_{\sigma= {\langle \sigma \rangle}}=0
\end{equation}
which, as expected, gives the self consistent relation
\begin{equation}
{\langle \sigma \rangle} = -  {\langle \sigma \rangle} \frac{\lambda}{N_c}  {\rm tr}\int i\frac {d^4 p}{(2\pi)^4} \frac{1} {\left[ -p^2 + \langle \sigma \rangle^2 \right]}.
\label{gengap}
\end{equation}
Chiral symmetry is broken when the true ground state lies in $\langle \sigma
\rangle \ne 0$. Therefore $\langle \sigma \rangle$ is the order parameter
which signals CSB which comes as no surprise since, by the equations of motion,
$\langle \sigma \rangle=  - 2G \langle \bar \psi_f \psi_f \rangle$ and the existence of a non vanishing quark condensate breaks CS. Equation (\ref {gengap}) also shows that $\langle \sigma \rangle$ is identical to the quark self energy within our approximation which allows us to write the effective quark mass as $M=m_c+\langle \sigma \rangle$. Finally, in the present work, the relevant quantity is the thermodymic potential, $\Omega_f$, which is defined as Landau's free energy at its minimum $\Omega_f={\cal F}_f(\langle \sigma \rangle)$. The equation of state can be obtained by using $\Omega_f=-P_f$. Then, Eq. (\ref{gentrace}) allows us to write the quark pressure, away from the chiral limit, as

\begin{equation}
P_f= -\frac {(M-m_c)^2}{4G} - \frac{i}{2}  {\rm tr}  \int  \frac {d^4 p}{(2\pi)^4} \ln \left[ -p^2 + M^2 \right ]
 \,\,.
 \label{presstrace}
\end{equation}
In order to obtain results valid at finite $T$ and $\mu$ in the presence of an external magnetic field $B$
one can use the following replacements, which come from dispersion relations for quarks \cite{eduana}

\[
p_0 \to i(\omega_\nu - i \mu_f)\,\,\,,
\]
\[
{\bf p}^2 \to p_z^2 +(2n+1-s) \;\;\;\;,{\rm with}\;\;\;\;\;s=\pm 1\,\, \;\;,\;n=0,1,\dots
\]
\[
\int \frac{d^4 p}{(2\pi)^4} \to i \frac{T |q_f| B}{2\pi} \sum_{\nu = -\infty}^{\infty} \sum_{n=0}^{\infty} \int \frac{d p_z}{(2\pi)} \,\,\,\,.
\]
In the above relations, $\omega_\nu= (2 \nu+1)\pi T$, with
$\nu=0,\pm 1,\pm 2,\ldots$ representing the Matsubara frequencies
for fermions while $n$ represents the Landau levels (LL) and $s$
represents the spin states which, at $B \ne 0$, must be treated
separately. As explained in Ref. \cite{chakra} one may understand
the origin of those replacements,  by recalling that, quantum
mechanically, the energy associated with the circular motion in the
$x-y$ plane is quantized in units of $2 q B$ due to the field in the
$z$ direction while the energy associated to linear motion along
$z$ is taken as  a continuous. All these levels
for which the values of $p_x^2+p_y^2$ lie between $2qBn$ and
$2qB(n+1)$
 now coalesce together into a single level characterized by $n$ whose number is given by
\begin{equation}
\frac{S}{(2\pi)^2} \int \int dp_x dp_y = \frac{SqB}{2\pi} \,\,,
\end{equation}
where $S$ is the area in the $x-y$ plane and $q$ stands for $|q_f|$.
Then, using those replacements, taking the trace, and summing over Matsubara's frequencies (see Appendix A) one obtains
\begin{eqnarray}
P_f&=& -\frac {(M-m_c)^2}{4G} + \frac{N_c }{2\pi} \sum_{s,n,f} (|q_f|B)\int \frac{d p_z}{(2\pi)}   E_p(B)  \nonumber \\
&+& \frac{N_c }{2\pi} \sum_{s,n,f} (|q_f|B)\int \frac{d p_z}{(2\pi)} \left \{ T \ln[ 1+ e^{-[E_p(B)+\mu_f]/T}] \right . \nonumber \\
&+& \left . T \ln [1+ e^{-[E_p(B)-\mu_f]/T}] \right \} \,\,\,,
\label{BTmu}
\end{eqnarray}
where $E_p(B)=\sqrt{p_z^2+(2n+1-s)|q_f|B + M^2}$. Next, by analyzing the degeneracy of  the lowest Landau level (LLL)  one can define $E_{p,k}(B)=\sqrt{p_z^2+2k|q_f|B + M^2}$ replacing $n$ by $k$ in the sum which appears in Eq. (\ref{BTmu}) which, now, also runs over the degeneracy label, $\alpha_k= 2-\delta_{k 0}$.
Being mainly concerned with the case $T=0,\mu \ne 0$, and $B \ne 0$ we  can take the limit $T \to 0$ in Eq. (\ref {BTmu}) arriving at (see Appendix A)
$$P_f(\mu,B)= -\frac {(M-m_c)^2}{4G} + P^{med}_f$$
\begin{equation}
+ \frac{N_c }{2\pi} \sum_{f=u}^{d} \sum_{k=0}^{\infty}\alpha_k (|q_f|B)
\int_{-\infty}^{\infty} \frac{d p_z}{(2\pi)}E_{p,k}(B),
\label{pressBmu}
\end{equation}
where the contribution from the medium is
\begin{equation}
P_f^{med}=\frac{N_c }{2\pi} \sum_{f=u}^{d} \sum_{k=0}^{\infty} \alpha_k (|q_f|B)\int_{-\infty}^{\infty} \frac{d p_z}{(2\pi)}[\mu_f -
E_{p,k}(B)]
 \,\,.
\end{equation}
Note that, although not explicitly written, a $\theta$ function (more specifically, $\theta(\mu_f -E_{p,k})$) must be considered as multiplying all $\mu_f$ dependent terms, such as $P_f^{med}$, appearing in our work (see Appendix A).  Simple power counting reveals  that the last term in Eq. (\ref {pressBmu}) is (ultra violet)
divergent while the second term, which
contains the in medium contributions, is finite since it has a
natural cut off given by the Fermi momentum, $p_{f,F}^2= \mu_f^2 -
M^2$. In Appendix B we show how  Eq. (\ref {pressBmu}) can acquire a physically more  appealing form by separating the (divergent) vacuum contribution from the (finite) magnetic field contribution. As a byproduct, those manipulations also produce more elegant relations in which the infinite sum over Landau levels appearing in the last term of Eq. (\ref {pressBmu}) are shuffled into Riemann-Hurwitz $\zeta$ functions.
 We finally get
\begin{equation}
P_f(\mu_f,B)= -\frac {(M-m_c)^2}{4G} +\left [P^{vac}_f+P^{mag}_f + P^{med}_f \right ]_{M}\,\,,
\label{pressBmu2}
\end{equation}
where the vacuum contribution reads
\begin{equation}
P^{vac}_f=2 N_c N_f  \int\frac{d^3 {\bf p}}{(2\pi)^3}   E_{\bf p}\,\,,
\end{equation}
with $E_{\bf p}= \sqrt {{\bf p}^2+ M^2}$ and $N_f=2$. We are now in position to present the explicit expressions for $P^{vac}_f,P^{mag}_f$, and $P^{med}_f$.
Let us start with the vacuum which, upon using a sharp non covariant cut off, $\Lambda$, can be written as
\begin{equation}
P^{vac}_f=- \frac{N_c N_f}{8\pi^2} \left \{ M^4 \ln \left [
    \frac{(\Lambda+ \epsilon_\Lambda)}{M} \right ]
 - \epsilon_\Lambda \, \Lambda\left[\Lambda^2 +  \epsilon_\Lambda^2 \right ] \right \},
\end{equation}
where we have defined $\epsilon_\Lambda=\sqrt{\Lambda^2 + M^2}$. The evaluations performed in Appendix B also give the following finite  magnetic contribution
\begin{eqnarray}
P^{mag}_f&=&\sum_{f=u}^d \frac {N_c (|q_f| B)^2}{2 \pi^2} \left \{ \zeta^\prime[-1,x_f]  \right . \nonumber \\
&-& \left . \frac {1}{2}[ x_f^2 - x_f] \ln x_f +\frac {x_f^2}{4} \right \}\,\,,
\end{eqnarray}
where   $x_f = M^2/(2 |q_f| B)$ while
$\zeta^\prime(-1,x_f)= d \zeta(z,x_f)/dz|_{z=-1}$ where $\zeta(z,x_f)$ is the Riemann-Hurwitz zeta function \cite {wolfram}. Finally, after integration, the medium contribution can be written as
\begin{eqnarray}
P^{med}_f&=&\sum_{f=u}^d \sum_{k=0}^{k_{f,max}} \alpha_k\frac {|q_f| B N_c }{4 \pi^2}  \left \{ \mu_f \sqrt{\mu_f^2 - s_f(k,B)^2} \right .\nonumber \\
&-& \left . s_f(k,B)^2 \ln \left [ \frac { \mu_f +\sqrt{\mu_f^2 -
s_f(k,B)^2}} {s_f(k,B)} \right ] \right \} ,
\label{PmuB}
\end{eqnarray}
where  $s_f(k,B)
= \sqrt {M^2 + 2 |q_f| B k}$.
The  upper Landau level (or the nearest integer) is defined by
\begin{equation}
k_{f, max} = \frac {\mu_f^2 -M^2}{2 |q_f|B}= \frac{p_{f,F}^2}{2|q_f|B}.
\label{landaulevels}
\end{equation}

Finally, the term $M$ entering the quark pressure is just the effective
self consistent mass at finite density and in the presence of an external magnetic field:

\begin{eqnarray}
M &=& m_c +\frac{ M G N_c N_f}{\pi^2} \left \{
\Lambda
\sqrt {\Lambda^2 + M^2} \right . \nonumber \\
&-& \left . \frac{M^2}{2}
\ln \left [ \frac{(\Lambda+ \sqrt {\Lambda^2 + {M^2}})^2}{{M}^2} \right ] \right \} \nonumber \\
&+&\sum_{f=u}^d \frac{ M |q_f| B N_c G}{\pi^2}\left \{ \ln\{\Gamma[x_f]\}  \right . \nonumber \\
&-& \left .\frac {1}{2} \ln (2\pi) +x_f -\frac{1}{2} \left [ 2 x_f-1 \right ]\ln[x_f] \right \} \nonumber \\
&-&\sum_{f=u}^d\sum_{k=0}^{k_{f,max}} \alpha_k \frac{ M |q_f| B N_c G}{\pi^2} \nonumber \\
&\times& \left \{\ln \left [ \frac { \mu_ f +\sqrt{\mu_f^2 -
s_f(k,B)^2}} {s_f(k,B)} \right ] \right \}\,\,,
\label{MmuB}
\end{eqnarray}
where we have used some $\zeta(z,a)$ properties given in Appendix B.
Note that our Eq. (\ref {pressBmu2}), although obtained in a different fashion, exactly agrees \footnote{When one sets $\mu_u=\mu_d=\mu$.} with the result obtained in Refs. \cite {klimenko03,klimenko} where the authors have used Schwinger's proper time formalism \cite {schwinger}.
In order to make this paper self contained  let us review, in the next section, some standard results obtained for symmetric quark matter.

\section{EoS for symmetric matter}

In this section we use  Eq. (\ref {pressBmu2}) to reproduce some of the most
important results concerning symmetric quark matter at $\mu \ne 0$ and/or $B
\ne 0$.  The result for this particular case can be readily obtained by
setting $\mu_u=\mu_d=\mu$.
It is worth emphasizing that, when the magnetic field is
turned on, the $u$ and $d$ quark densities are not the same, as it is discussed
next. In this sense, both quark chemical potentials are identical, but
the densities are not and hence, matter is not strictly symmetric.
Besides this, we have opted to keep the nomenclature.

\subsection{ EoS for symmetric  matter at $\mu \ne 0$ and $B=0$}

This case has been extensively discussed in the literature and we refer the
reader to Refs. \cite{klevansky,buballa} for more details.
 
By setting $B=0$ one obtains the following relation for the pressure

\begin{eqnarray}
P_f (\mu,0)&=& -\frac{[M_\mu-m_c]^2}{4 G} \nonumber \\
&-& \frac{N_c N_f}{8\pi^2} \left \{ M_\mu^4 \ln \left [
    \frac{(\Lambda+ \epsilon_\Lambda)}{M_\mu} \right ]
- \epsilon_\Lambda \, \Lambda\left[\Lambda^2 +  \epsilon_\Lambda^2 \right ] \right \} \nonumber \\
&+&\frac{N_c N_f}{8 \pi^2} \left \{ M_\mu^4 \ln \left [ \frac{(\mu+
      p_F)}{M_\mu} \right ]
+ \frac{5\mu}{3}  p_F^3 - \mu^3\ p_F \right\}, \nonumber
\label{pressuremu}
\end{eqnarray}
where $p_F^2= \mu^2-M_\mu^2$. The effective mass, $M_\mu$, satisfies
\begin{eqnarray}
M_\mu &=& m_c +\frac{ M_\mu\, G\, N_c\, N_f}{\pi^2} \left \{ \Lambda
\epsilon_\Lambda - {M_\mu^2}
\ln \left [ \frac{\Lambda+ \epsilon_\Lambda}{M_\mu} \right ] \right. \nonumber \\
&-& \mu \, p_F +\left .
{M_\mu}^2
\ln \left [ \frac{\mu+ p_F}{{M_\mu}} \right ] \right \}.
\end{eqnarray}
The normalized pressure is simply given by
$P_{f, {\rm eff}}(\mu,0) = P_f(\mu,0)|_{M_\mu} - P_f(0,0)|_{M_0}$ where
\begin{eqnarray}
 P_f (0,0)&=&-\frac{ [M_0-m_c]^2}{4 G}
-\frac{N_c N_f}{8 \pi^2} \left \{ M_0^4 \ln \left [ \frac{\Lambda+
      \epsilon_{\Lambda,0}}{M_0} \right ]
 \right . \nonumber \\&-& \left .
  \Lambda\, \epsilon_{\Lambda,0}( \Lambda^2 + \epsilon_{\Lambda,0}^2) \right \},
\label{Pzero}
\end{eqnarray}
with $\epsilon_{\Lambda,0}=\sqrt{M_0^2+ \Lambda^2}$.
The effective quark mass, $M_0$, appearing in Eq. (\ref{Pzero}) is just $M_\mu$
evaluated at $\mu=0$. Namely
\begin{equation}
M_0 =m_c +\frac{ M_0\, G\, N_c\, N_f}{\pi^2} \left \{ \Lambda
\epsilon_{\Lambda,0} - {M_0^2}
\ln \left [ \frac{\Lambda + \epsilon_{\Lambda,0} }{M_0} \right ]\right \}.
\end{equation}

Finally, at $B=0$ and $\mu \ne 0$, the equation of state for the
quarks is given by ${\cal E}_f (\mu,0) = -P_{f, {\rm eff}}(\mu,0)+
\mu \rho (\mu,0)$ where $\rho(\mu,0) = d P_f(\mu ,0)/d \mu$ is
the mean field result
\begin{equation}
\rho(\mu,0)= \frac{N_c N_f}{3\pi^2} [\mu^2 - M_\mu^2]^{(3/2)} = \frac{N_c N_f}{3\pi^2}p_F^3 \,\,.
\end{equation}
 We can also
define the {\it bag constant}
\begin{equation}
{\cal B} = P_{f,{\rm eff}}(\mu,0)|_{m_c} -P_{f,{\rm eff}}(\mu,0)|_{M_\mu} \,\,.
\end{equation}
As emphasized in Ref. \cite {buballa} one should remark that, in the same way as in the bag model, $\cal B$ describes the pressure
difference between the trivial and non trivial vacuum, but its is not an input of the model, being a dynamical
consequence of the interactions which leads to vacuum masses $M(0,0) \ne m_c$. Using our chosen values for $G$ and
$\Lambda$ one obtains ${\cal B}=$(181 MeV)$^4$. 
Regarding CSB, one obtains the quark effective mass, $M \simeq 400 \, {\rm
  MeV}$ for $\mu=0$ and  observes a first order transition for chiral
symmetric matter at $\mu_c \simeq 360 {\rm MeV}$.
A quantity of particular interest for the present work is the energy per baryon as a function of the density. The relations obtained above for the $B=0$ case will allow us to discuss, at the end of this section, the influence of the magnetic field regarding the stability of quark matter.

\subsection{EoS for symmetric matter at $\mu =0$ and $B \ne 0$.}

In this subsection we set $\mu=0$ and concentrate in the behavior of the quark
condensate under the influence of an external magnetic field. One of the most
remarkable effects of this external field regards its role so as to enhance
chiral symmetry breaking. This issue is related to the phenomenon known as
magnetic catalysis which has been well exploited by Klimenko and collaborators
among others \cite{klimenko,klimenko03}.
Going back to our general relation for the pressure, Eq. (\ref {pressBmu2}),
and setting $\mu_u=\mu_d=\mu=0$ one obtains that the relevant pressure
is given by
\begin{equation}
P_f(0,B)= -\frac {[M_B-m_c]^2}{4G} +\left [P^{vac}_f+P^{mag}_f\right]_{M_B} \,\,,
\label{pressBzero}
\end{equation}
with the effective quarks mass at $\mu =0$ and $B \ne 0$, $M_B$, being determined by
\begin{eqnarray}
&&M_B = m_c +\frac{ M_B G N_c N_f}{\pi^2} \left \{
\Lambda
\sqrt {\Lambda^2 + M_B^2} \right . \nonumber \\
&&- \left . \frac{M_B^2}{2}
\ln \left [ \frac{(\Lambda+ \sqrt {\Lambda^2 + {M_B^2}})^2}{{M_B}^2} \right ] \right \} \nonumber \\
&&+\sum_{f=u}^d \frac{ M_B |q_f| B N_c G}{\pi^2}\left \{ \ln\ \Gamma[x_f] - \frac {1}{2} \ln (2\pi) \right . \nonumber \\
&& \left .+ x_f -\frac{1}{2} \left [ 2 x_f-1 \right
]\ln x_f \right \}
\label{MmuzeroB}
\end{eqnarray}
where now $x_f=M_B^2/(2|q_f|B)$ (see Eq.(\ref{MmuB})).

\begin{figure}[h]
\begin{center}
\includegraphics[width=7.cm]{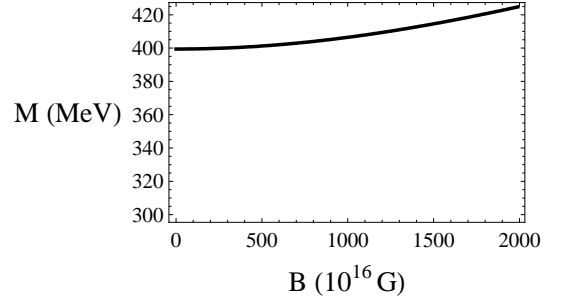}
\caption{Quark effective mass as a function of $B$ showing how the
latter enhances CSB (magnetic catalysis).} \label{marcusfig1}
\end{center}
\end{figure}
Fig. \ref{marcusfig1} shows the behavior of the effective quark mass
as a function of B indicating that the former increases (although
slightly) with the latter which stabilizes the condensate so that
the gap equation always has a non trivial solution for finite $B$.
As pointed out in Ref. \cite{klevansky} the $B$ field facilitates
the binding by antialigning the helicities of the quark and the
antiquark, which are then bound by the NJL interaction while an
electric field, $E$, has an opposite effect opposing condensate
formation by polarizing the ${\bar \psi}_f \psi_f$ pairs.

\subsection{EoS for symmetric matter at $\mu \ne 0$ and $B \ne 0$.}

\begin{figure}[h]
\begin{center}
\includegraphics[width=7.cm]{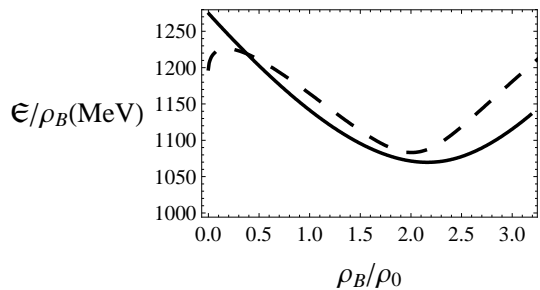}
\caption{ Energy per baryon as a function of the baryonic density
for symmetric quark matter. Continuous line $B = 2 \times 10^{19}$G, dashed line
$B=0$. 
} \label{marcusfig2}
\end{center}
\end{figure}

In this subsection, still considering only the symmetric case, we shall review how $B$
stabilizes quark matter due to magnetic catalysis,  according to Ref. \cite {klimenko03}.
For this we can consider the general relation, Eq. (\ref {pressBmu2}) with $\mu_u=\mu_d=\mu$,
from which the quark density is easily extracted as
\begin{equation}
\rho(\mu,B) = \sum_{f=u}^{d} \sum_{k=0}^{k_{f, max}}\alpha_k \frac{
|q_f| B N_c }{6 \pi^2}   k_{F,f}(k,s_f) \,\,, \label{rhoQmuB}
\end{equation}
where $k_{F,f}(k,s_f) =\sqrt{\mu^2 - s_f(k,B)^2}$ while the effective pressure is defined by
\begin{equation}
P_{f ,{\rm eff}}(\mu,B) = P_f(\mu,B)|_{M}-P_f(0,B)|_{M_B} \,\,.
\end{equation}
Then, the energy density follows as
\begin{equation}
{\cal E}_{ \beta f} (\mu,B)= -P_{\beta f,{\rm eff}}(\mu,B) + \mu
\rho(\mu,B).
\end{equation}
In all the above equations $M$ is given by Eq. (\ref {MmuB}) with the obvious
substitution $\mu_u=\mu_d=\mu$.  In Fig. \ref{marcusfig2} we present some results
which compare  the energy
per baryon (${\cal E}/\rho_B$) as a function of the  density for $B=0$ and $B= 2\times 10^{19} \, $G.

It is shown that for $B=0$ the curve has an absolute minimum at
$\rho_B \simeq 2 \rho_0$ and that at $\rho_B=0$ there is a
local minimum corresponding to the real empty physical QCD-vacuum.
This fact indicates that, in this case, the vacuum is metastable but
the situation changes drastically in the presence of an external
magnetic field and the graph has a maximum at $\rho_B=0$ so that at
arbitrarily small densities stable quark droplets appear in the
system. One then concludes that the external magnetic field supports
the creation of stable quark matter. Later, when discussing our
numerical results, in section VI, we again address some issues
related to the symmetric case.
As one can see, in  figure \ref{marcusfig2}, the minima obtained for both cases have
different values. The case of symmetric magnetized quark matter has a  minimum
at a density lower
than the  one  obtained for symmetric quark matter. The concavity of the curve
in that case is also greatest.

Now, after reviewing some of the most important issues regarding how symmetric quark matter is affected by
the presence of an external magnetic field we can
incorporate $\beta$-equilibrium in our results to consider 
stellar matter.

\section {Asymmetric Quark Matter with $\beta$-equilibrium in a magnetic field}

In a star with quark matter we must impose both, $\beta$ equilibrium and
charge neutrality \cite{Glen00}. Through out this paper we only
consider the latest stage in the star evolution, when entropy is
maximum and neutrinos have already diffused out. The neutrino
chemical potential is then set to zero.  For $\beta$-equilibrium
matter we must add the contribution of  leptons (electrons and muons) in a magnetic field to the energy
density and pressure. The relations between the chemical potentials
of the different particles are given by
\begin{equation}
\mu_d=\mu_u+\mu_e, \qquad  \mu_e=\mu_\mu.
\label{qch}
\end{equation}
For charge neutrality we must impose
\begin{equation}
\rho_e+\rho_\mu=\frac{1}{3}(2\rho_u-\rho_d).
\label{chneutrality}
\end{equation}
Now, Eqs. (\ref {qch}) and (\ref {chneutrality}) will have to be satisfied together with the self consistent
relations for the quark effective mass during  the numerical evaluations. 
The EoS
for the leptonic sector is also needed. Let us now obtain this EoS by recalling that,
as emphasized in the introduction, the total leptonic pressure, $
P_l(\mu_l,B)$, is quickly recovered from the quark pressure
, $P_f(\mu_f,B)$, upon performing obvious replacements such as $f\to l$, and $N_c=1$. Also, because the leptonic
sector does not have the analog of the quartic interaction, $G\to 0$, so that when translating the results one
takes $M\to m_c \to m_l$  which lead to
\begin{equation}
P_l(\mu_l,B)= \left [P^{vac}_l+P^{mag}_l + P^{med}_l \right ]_{m_l}\,\,.
\label{pressBmulep}
\end{equation}
Being evaluated at the same (one loop) approximation level all terms have exactly the same mathematical structure as those in $P_f(\mu_f,B)$ (including the divergences in the vacuum contribution). A major difference is that all terms in Eq. (\ref {pressBmu}) are evaluated with the bare $m_l$ reflecting the fact that an undressed lepton propagator has been used. At this one loop level of approximation the leptonic contribution is that of a free gas of relativistic fermions and the divergences contained in the vacuum contribution can be properly absorbed with a zero point subtraction. However, note that according to our normalization procedure we require
$P_l(\mu_l,B)=0$ at $\mu_l=0$ as in the quark case which leads to the following effective pressure for the leptonic sector
\begin{equation}
P_{l,{\rm eff}}(\mu_l,B)= \left [P_l(\mu_l,B)-P_l(0,B) \right ]_{m_l}=P^{med}_l \,\,,
\end{equation}
since $P_l(0,B)= P^{vac}_l+P^{mag}_l$ and all these quantities are written in terms of the $\mu_l$-independent mass, $m_l$.
The result  shows that, at the one loop level, only the following (finite) medium contribution has to be considered
\begin{eqnarray}
P_{l,{\rm eff}}(\mu_l,B)&=&\sum_{l=e}^\mu \sum_{k=0}^{k_{l,max}} \alpha_k\frac {|q_l| B }{4 \pi^2}   \left \{ \mu_l \sqrt{\mu_l^2 - s_l(k,B)^2} \right .\nonumber \\
&-& \left . s_l(k,B)^2 \ln \left [ \frac { \mu_l +\sqrt{\mu_l^2 - s_l(k,B)^2}}
    {s_l(k,B)} \right ] \right \},\nonumber \\
&&
\end{eqnarray}
Then, the leptonic density is easily evaluated yielding
\begin{equation}
\rho_l(\mu_l,B) = \sum_{l=e}^\mu \sum_{k=0}^{k_{l, max}}\alpha_k \frac{ |q_l| B }{2 \pi^2}   k_{F,l}(k,s_l) \,\,,
\label{rholmuB}
\end{equation}
where $k_{F,l}(k,s_l) =\sqrt{\mu_l^2 - s_l(k,B)^2}$.
Finally, the leptonic energy density reads
\begin{equation}
{\cal E}_{l} (\mu_l,B)= -P_{l ,{\rm eff}}(\mu_l,B) + \mu_l \rho_l (\mu_l,B)\,\,\,,
\end{equation}
where, again, a sum over the repeated ($l$) indices is implied.
Finally, the total effective pressure corresponding to the theory described by
${\cal L}_{\beta f}$ in the presence of a constant external magnetic
field, $B$, is
\begin{equation}
P_{\beta f, {\rm eff}}(\mu_f,\mu_l,B) = P_{f,{\rm eff}}(\mu_f,B) + P_{l,{\rm eff}}(\mu_l,B) \,\,\,.
\end{equation}

We now have all the ingredients to evaluate the EoS in $\beta$ equilibrium since
\begin{equation}
{\cal E}_{ \beta f} (\mu_f,\mu_l,B)= -P_{\beta f ,{\rm
eff}}(\mu_f,\mu_l,B) + \mu_l \rho_l + \mu_f \rho_f\,\,\,,
\end{equation}
where the quark and leptonic densities are given by Eqs. (\ref {rhoQmuB}) and (\ref {rholmuB}) respectively.

\section{Results and Discussion}

We start by discussing, in more detail, the effects of the magnetic field on
the EoS of symmetric matter described
by the usual SU(2) version of the NJL model without $\beta$-equilibrium
whose inclusion will be addressed  afterwards.
Whenever mentioned in the figures, $B_0=10^{19}$ G.
By symmetric matter we  mean $ud$ matter for which the chemical potential of both
particles are equal. For a zero magnetic field this is equivalent to having symmetric
matter. For finite magnetic fields due to the different electrical charge of both quarks and
the appearance of Landau levels, $ud$ is symmetric only for  restricted  densities.

\begin{figure}[h]
\begin{tabular}{ccc}
\end{tabular}
\end{figure}
\begin{figure}[h]
\begin{tabular}{ccc}
\includegraphics[width=7.cm]{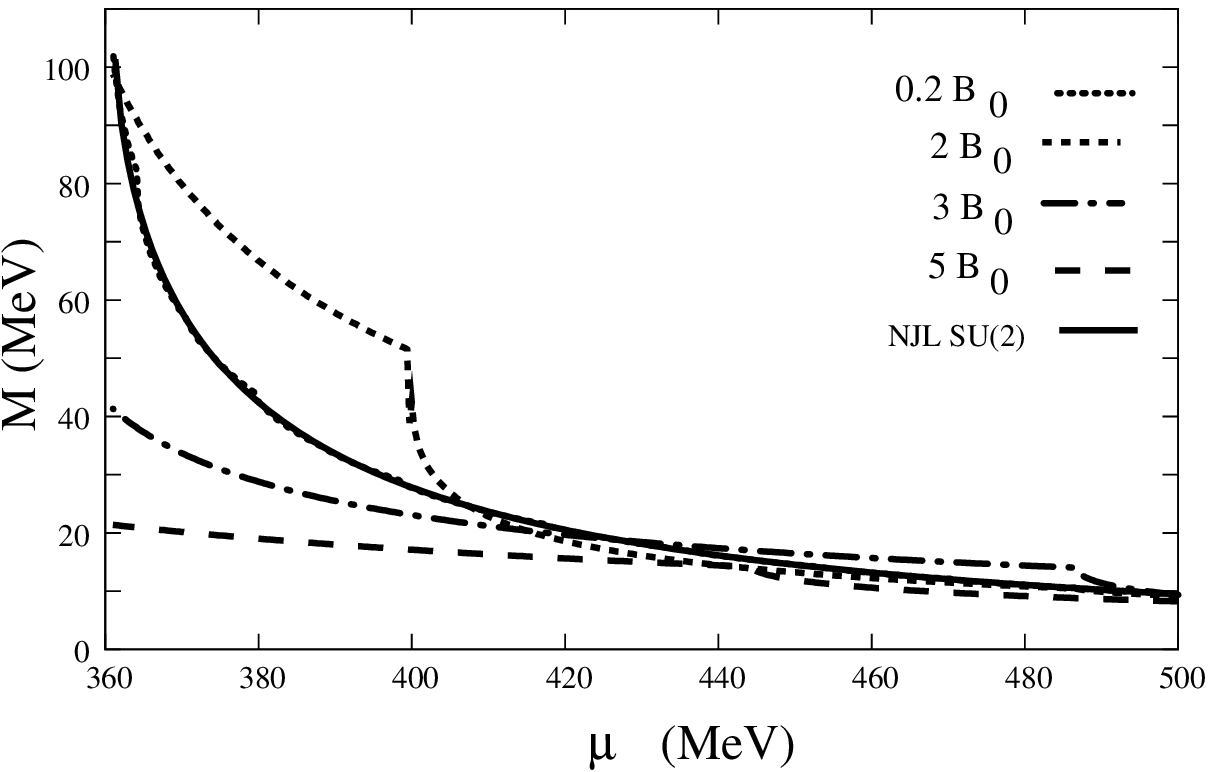}\\
\includegraphics[width=7.cm]{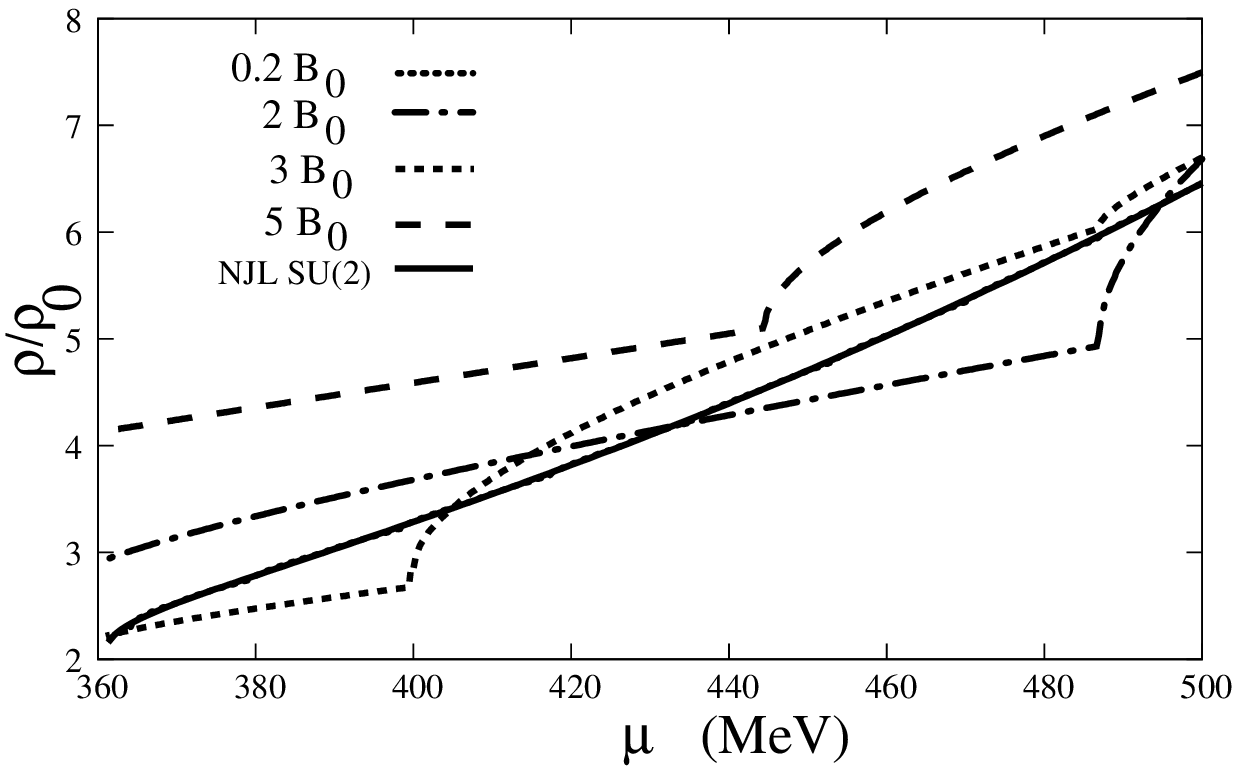}  
\end{tabular}
\caption{a) Effective quark mass and b) baryonic density as a
function of the chemical potential for matter subject to different
values of the magnetic field. $\rho_0$ was chosen as 0.17 fm$^{-3}$
and $B_0=10^{19}$ G.}
\label{figsnovas}
\end{figure}
\begin{figure}[h]
\begin{center}
\begin{tabular}{c}
\includegraphics[width=7.cm]{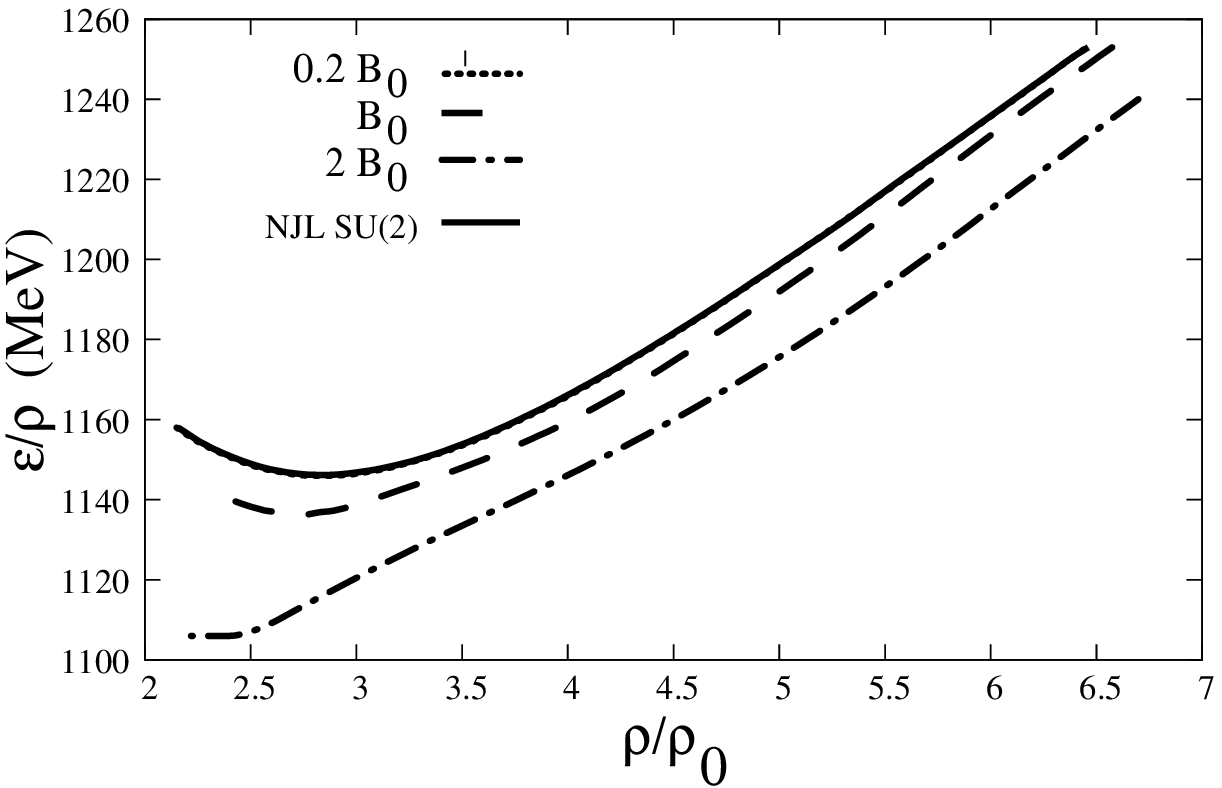} \\
\includegraphics[width=7.cm]{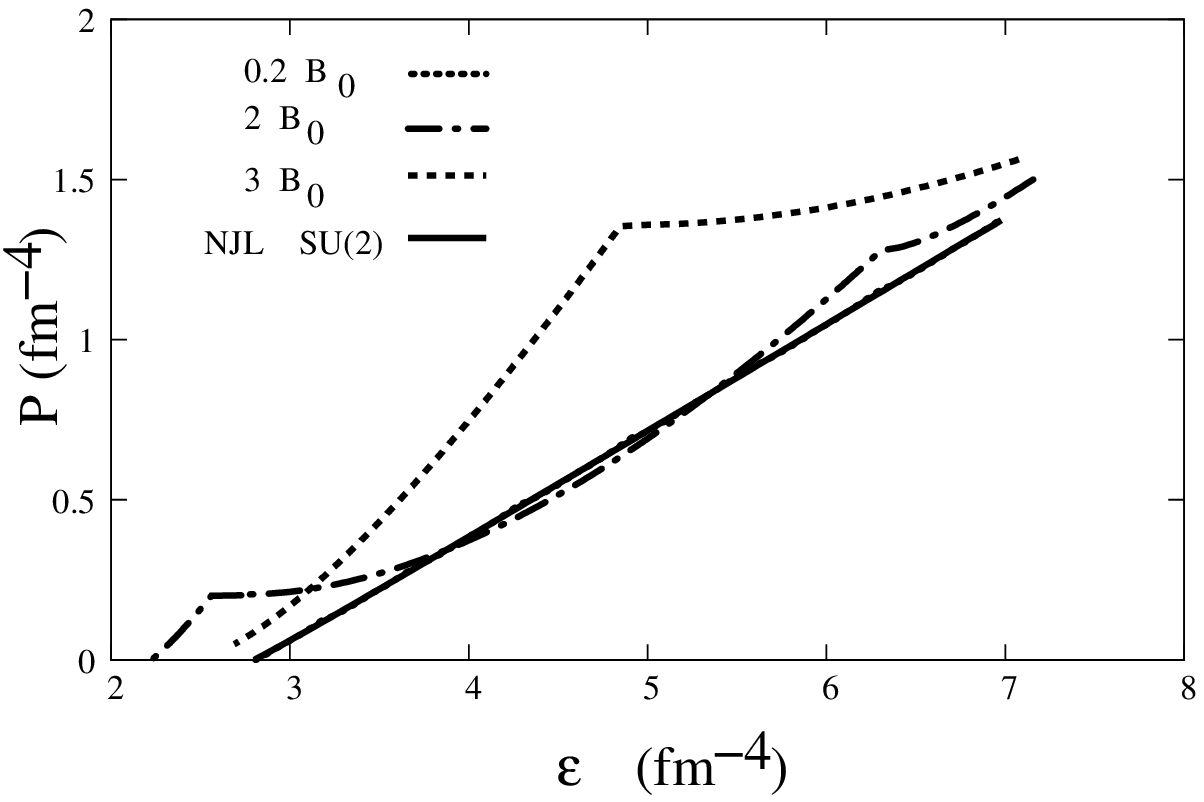}\\
\end{tabular}
\end{center}
\caption{ a) Binding energy and b) EoS for different values of the
magnetic field for $ud$ matter with equal chemical potentials. $B_0=10^{19}$ G.}
\label{figsnovas2}
\end{figure}

In the next four figures, i.e., Fig.  \ref{figsnovas}a,\ref{figsnovas}b,
\ref{figsnovas2}a and \ref{figsnovas2}b, the results for 
$B=0.2 \times 10^{19}$ G are almost coincident with the results for 
magnetic free matter. They are kept so that minor differences can be seen.

In figures \ref{figsnovas}a and \ref{figsnovas}b we compare the
effective quark mass and the baryonic density as a function of the
chemical potential for matter subject to different values of the
magnetic field. One can see that a magnetic field of the order of $0.2
\times 10^{19}\,\,{\rm G}$ barely affects the effective mass as compared
with the results for ordinary matter (not subject to the magnetic
field).
Due to the Landau quantization, the increase of the strength of the
magnetic field provokes a decrease of the number of the filled
LL and the amplitude of the oscillations is more clear in the
graphics. For each value of the magnetic field, the kink
appearing at the smallest chemical potential corresponds to the case when
only the first LL has been occupied.
For $B=5\times 10^{19}$ G matter is totally polarized for chemical
potentials below 490 MeV. For the small values of the magnetic
fields the number of filled LL is quite large and the effects of the
quantization are less visible.
For the larger magnetic fields the chiral symmetry restoration
occurs for smaller values of the chemical potentials which, however,
correspond to larger densities as can be seen from
Fig. \ref{figsnovas}b).

In Fig. \ref{figsnovas2}a one can see that the inclusion of the
magnetic field makes matter more and more bound.
The energy per baryon $E/A$ of magnetized quark matter
described by the SU(2) version of the NJL model is less bound than nuclear
matter made of iron nuclei, $\frac{E}{A}|_{^{56}Fe} \sim 930$ MeV,  
which means that quark matter is
not the preferential ground state matter even in presence of the magnetic fields
under consideration \cite{buballa96,buballa99}.

In Fig. \ref{figsnovas2}b we plot the EoS for different values of
the magnetic field. Once again one can see that the magnetic fields
modify the EoS but the modifications are more significant when the
magnetic fields reach values higher than $0.2\times 10^{19}$G;
as can be seen these graphics show also oscillations every time that a
different number of LL is filled. As the magnetic field increases the number
of  LL for a given interval of energy is reduced, increasing the gap between
them and making more visible the oscillations, the so-called
Haas van Alphen effect \cite{haas}.

We have identified the densities at which pressure
is zero, for different values of the magnetic field. 
This   corresponds to the saturation density of
symmetric quark matter when $B=0$, or to
$\mu_u=\mu_d$ otherwise.
In particular, for $B=0.2 \times 10^{19}$ G, the density is 1.44
fm$^{-3}$ and for $B=2 \times 10^{19}$ G, the density is  1.19
fm$^{-3}$, and for $B=3 \times 10^{19}$ G, the density is around
2.7 fm$^{-3}$, i.e. the saturation density decreases as the magnetic
field increases for magnetic fields lower than  $B=2 \times 10^{19}$ G.
For larger fields the opposite may occur.
The  minimum of the binding energy is the result of two contributions with
different behavior: on one hand, due to
Landau quantization the kinetic energy decreases with the increase of the
magnetic field and on the other, the effective bag parameter defined by the
interaction terms minus the vacuum energy increases because the restoration
of chiral symmetry occurs at larger densities. The saturation density depends
on how fast each contribution changes with density.

\begin{figure}[h]
\begin{center}
\begin{tabular}{c}
\includegraphics[width=7.cm]{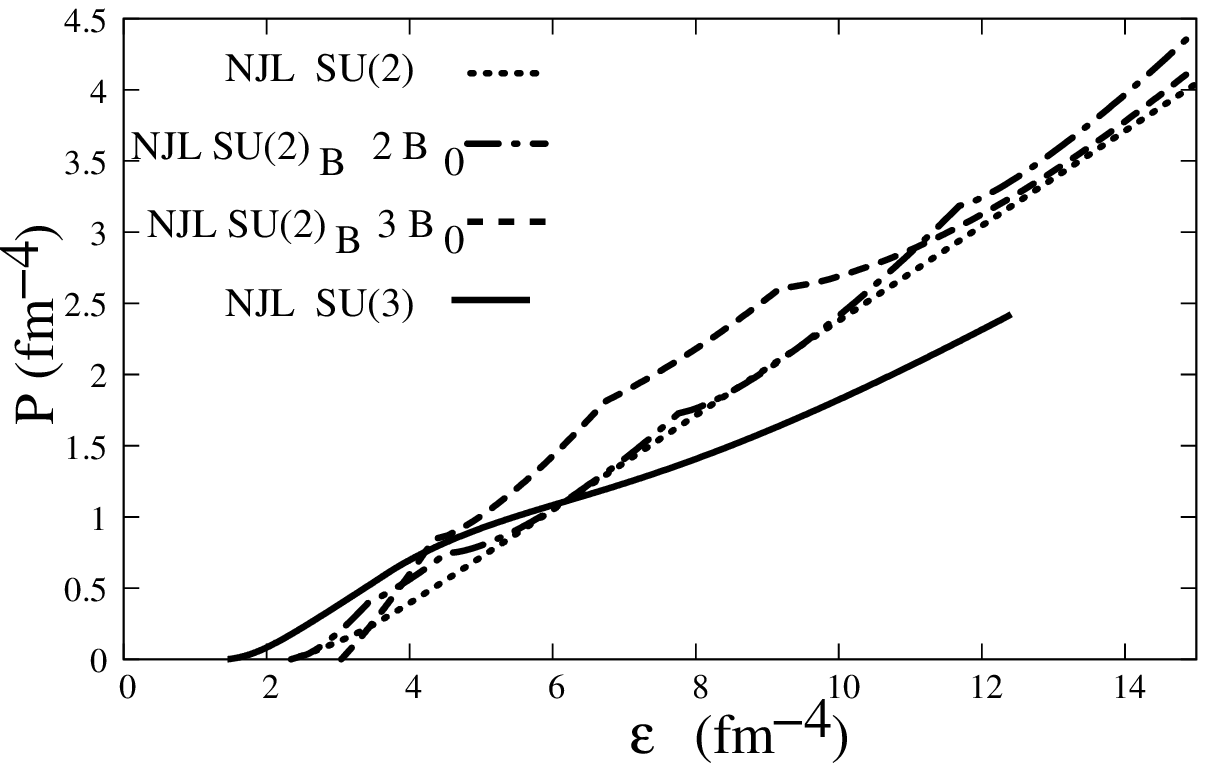} \\
\includegraphics[width=7.cm]{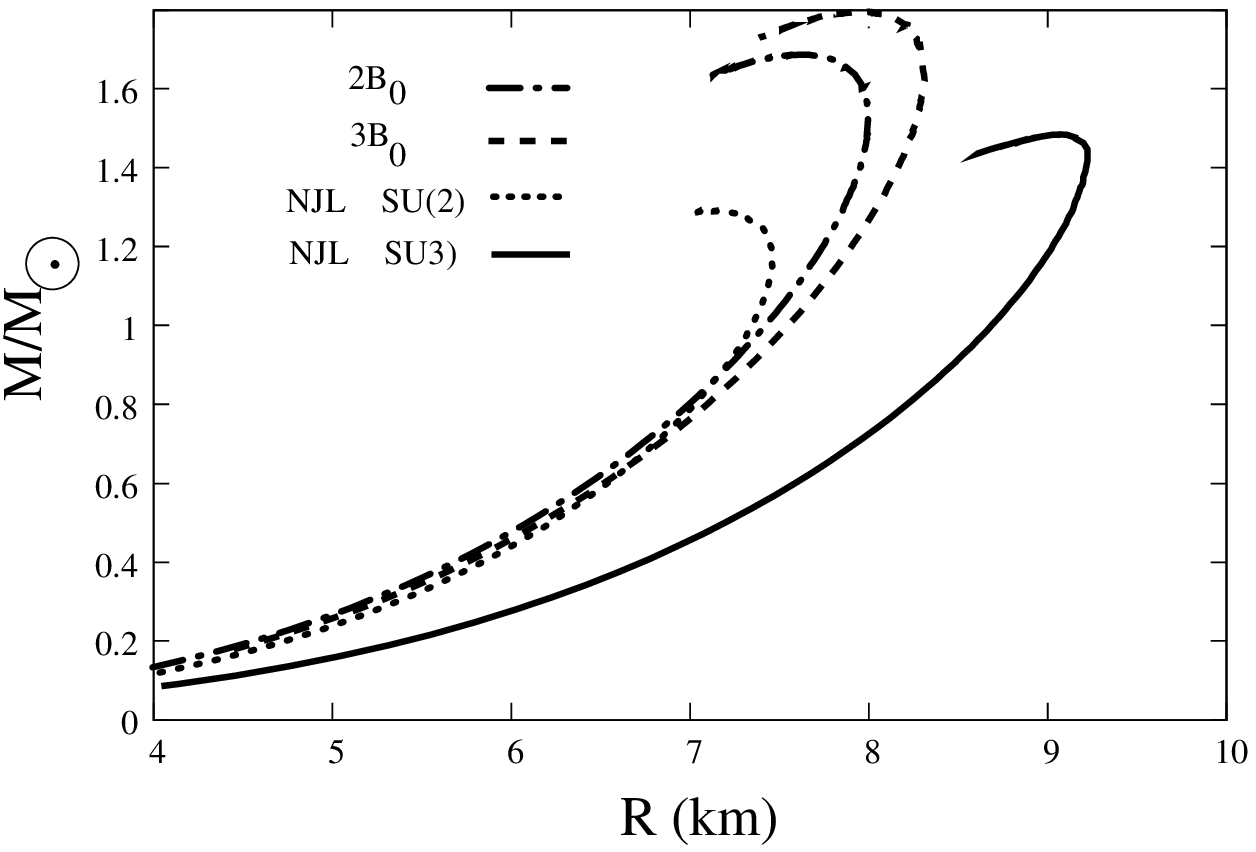}\\
\end{tabular}
\end{center}
\caption { a) EoS for quark matter in $\beta$-equilibrium and b)
resulting stellar properties for the SU(2) version
  of the NJL model both with and without magnetic field and the SU(3) version
  without magnetic field. In the figures, $B_0=10^{19}$ G.}
\label{figsnovas3}
\end{figure}

We now study quark matter in $\beta$-equilibrium, the original
motivation of our studies aimed at understanding the constitution of magnetars.
In what follows, NJL SU(3) refers to the EoS of quark matter with
  $u,\,d,\,s$ quarks and parameters used in
\cite{quarks} and NJL SU(2) refers to $ud$ quark matter. In both cases, matter is
not subject to a magnetic field. The designation (NJL SU(2))$_B$ refers to $ud$ quark matter
subject to a magnetic field. {The reader should keep
in mind that here the comparison with the SU(3) case has only a qualitative character  since both, the SU(2) and SU(3),  versions
employ different parametrization sets flawing any conclusion related to quantitative aspects.}  Just as in symmetric matter, also for matter in
$\beta$-equilibrium, the effects of Landau quantization are clearly
seen in the EoS (see Fig. \ref{figsnovas3}a ).

The strange quark matter within the MIT bag has been discussed in 
\cite{chakra,helena}.
Once the $s$-quarks are removed, so that only $u,d$ matter is considered one 
can see that although the overall behavior is similar, there are clear 
differences between both models.
The zero pressure density decreases with the increase of the magnetic
field for the MIT model while there is no clear trend for the NJL.
These densities are given in Table\ref{press0}, where only $u$ and $d$ quarks
are taken into account. For quark stars this density defines the density at the
surface. We should remember, however that it is only a field above
10$^{19}$ G which has noticeable effects and, at the surface we
expect much smaller fields.

As for hadronic stars, the density is zero for a null pressure at
the surface if no magnetic field is included. The magnetic field gives rise to
an increase of the density in the surface \cite{prakash,cp}. In this sense, the
NJL model predicts a more similar behavior, i.e., an increase of the density
at zero pressure as a result of the magnetic field, whilst the MIT model
predicts the opposite.

\begin{table}[h]
\begin{tabular}{cccccc}
\hline
$B$ (G) & 0 & $0.1 B_0$ & $B_0$ & $2 B_0$ & $3 B_0$\\
\hline
$\rho_{NJL}$ (fm$^{-3}$)& 0.40 & 0.47 & 0.48 & 0.44 & 0.54 \\
$\rho_{MIT}$(fm$^{-3}$) & 0.54 & 0.47 & 0.45 & 0.36 & 0.31 \\
\hline
\end{tabular}
\caption{Baryon densities of $\beta$-equilibrium matter for which the
pressure becomes negative, for the NJL model and the MIT model with the bag
pressure (180 MeV)$^4$. In both models only $u$ and $d$ quarks were considered.}
\label{press0}
\end{table}

The saturation density is defined by the way chiral symmetry is restored:
a chiral symmetry restoration at smaller energies implies a smaller
saturation density. As discussed before, the saturation in the NJL model
is a balance between the attractive scalar field which saturates at
chiral symmetry restoration and the kinetic term which gains importance once
chiral symmetry restoration has occurred.

\begin{figure}[h]
\begin{center}
\includegraphics[width=7.cm]{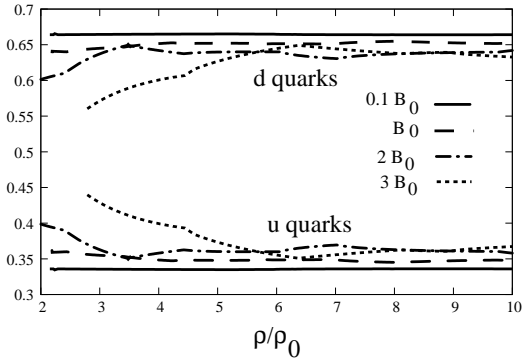}
\end{center}
\caption {Quark population ($\rho_i/\rho, i=d,u$ in terms of density for
  $B=0.1 B_0$ $B_0$, $2 B_0$ and $3 B_0$, with $B_0=10^{19}$ G respectively
  for d quarks from top to bottom and u quarks from bottom to top.}
\label{fraction}
\end{figure}

The conditions of charge neutrality and $\beta$-equilibrium give rise to
different chemical potentials, and therefore different quark
fractions, as seen in Fig. \ref{fraction}.  While for $B=0$
$\beta$-equilibrium matter is formed essentially by two thirds of $d$ quarks
and one third of $u$ quarks and only a very small fraction of electrons,
the presence of the magnetic field changes the particle fractions.
Matter becomes more symmetric for a finite magnetic field larger than
$\sim 10^{19}$ G, and a larger fraction of electrons and muons occurs.
In Fig. \ref{fraction} the lepton population is not shown so that the
effects
of the magnetic field on the quark population can be better noticed.

In Fig. \ref{figsnovas3}a we show the EoS for $\beta$-equilibrium quark stellar matter. We compare the $SU(2)$ and $SU(3)$ NJL EoS for magnetic free matter with the EoS for $B=2\times 10^{19}$ G and  $B=3\times 10^{19}$ G. In the last two EoS the effect of the LL quantization is clearly seen. Both EoS are harder than the corresponding $B=0$ EoS. 

Notice that, contrary to hadronic matter in
$\beta$-equilibrium, where the magnetic field makes the EoS softer
\cite{prakash,cp}, the NJL model predicts a slightly harder EoS, as seen in
Fig. \ref{figsnovas3}a.

To obtain the properties of the stars described by these EoS,  we have added
the contribution of the magnetic field to the pressure and to the energy
density, $B^2/2$. As discussed in section II the formal consideration  of this
term is related to the prescription adopted to normalize the pressure. At the
surface the magnetic field should be not larger than $\sim 10^{15}$ G and,
therefore, we have  introduced a density dependent magnetic field \cite{chakra97}
$$B(\rho)=B_{s}+B_i\, \left[1- exp[-\alpha (\rho/\rho_0)^\gamma]\right],$$ 
where $\rho_0$ is the saturation density, $B_s=10^{15}$ G is the magnetic
 field  at the surface, $B_i$ is the magnetic field  at the
interior for large densities, and the parameters $\alpha=5\times 10^{-5}$ and $\gamma=3$ were chosen in such a way 
that the field increases  fast with density to its central value but still  describes correctly the surface, namely with a zero  pressure.

 The properties of compact stars were obtained from the integration of the
Tolman-Oppenheimer-Volkoff equations,  using the EoS obtained with the density
dependent magnetic field and which includes the magnetic field contribution.
For  $B=3\times 10^{19}$ G the magnetic field contribution is as large as the
contribution of stellar matter. This is seen from the properties of the star
with maximum mass: the gravitational mass becomes larger than the baryonic
mass, (see Table II).

 The SU(3) EoS is also included. It becomes softer after
the onset of the $s$ quark, which occurs at a quite high density in
this model \cite{dp04}. The SU(3) version of the NJL model is obviously
more complete and the chiral symmetry restoration of the $s$ quark produces
a smooth change of declination in the EoS around 6 fm$^{-4}$ as can be seen
in Fig. \ref{figsnovas3}. As this energy density is very high,
the strangeness content of the star is small. The comparisons with
the SU(3) NJL version has to be considered with care because this version of
the model was fitted to a different set of variables, which, as we can see from
Fig. \ref{figsnovas3},  describe stellar matter (below 5 fm$^{-4}$)
in a different way from the SU(2) version.

The results of the integration of the TOV equations
are shown in Table II and
in figure \ref{figsnovas3}b. 
We do not take into account the anisotropy
introduced by the magnetic field \cite{Alain} and have assumed a spherical
configuration.
Compact stars with poloidal magnetic fields were studied in \cite{prak01}.

One can see that the results for the masses and radii of the maximum mass
stable configuration  obtained with the inclusion
of the magnetic field equal to $2 \times 10^{19}$ G and  $3 \times 10^{19}$ G  in the SU(2) version
of the NJL model is  larger than the corresponding SU(2)
magnetic free star.  The radius is still quite small,
below 8 km,
much smaller than the average neutron star value, $\sim 12-15$ km but the mass
is larger than the value
1.4 M$_\odot$ which many neutron stars have. As referred before, the gravitational mass for the two cases including the magnetic field
is larger than the baryonic mass due to the inclusion of the $B^2/2$ term in
the EoS.

\begin{table}[h]
\begin{tabular}{lccccccccc}
\hline
type & $B_i$ & $M_{\max}$ & $M_{b\max}$ &  $R$ & $\varepsilon_0$ & $B$\\
& Gauss & $(M_{\odot}$)&($M_{\odot}$)& (km) & (fm$^{-4}$) & ($10^{19}$ G)\\
\hline
NJL SU(3) \cite{quarks}& 0 &1.47 & 1.56 & 9.02 & 7.52 & 0 \\
NJL SU(2) & 0 & 1.29 & 1.25 & 7.11 & 13.5 & 0\\
(NJL SU(2))$_B$ & $2 \times 10^{19}$ & 1.69 & 1.55 & 7.63 & 12.0& 0.62\\
(NJL SU(2))$_B$ & $3 \times 10^{19}$ & 1.80 & 1.60 & 7.97 & 10.8& 0.63\\
\hline
\end{tabular}
\label{stars}
\caption{Quark star properties for the EoSs described in the text.}
\end{table}

\section{Conclusions}

The role of the magnetic field effects on quark matter has an
importance of its own and it has already been exploited in the literature.
One of the conclusions is that 
the consequences of Landau quantization are not negligible for large
values of the magnetic field.
Quark matter is generally described by the MIT bag model.
In this work we tackle the inclusion of the magnetic field
in quark matter described by the SU(2) version of the NJL model,
known to have more realist features.

Using the standard large-$N_c$ technique we have evaluated the effective
potential for the SU(2) version of the NJL model in 3+1 dimensions which has
allowed us to obtain its thermodynamical potential at finite density and in
the presence of an external magnetic field. Then, we have reviewed some of the
most important well known results related to three different situations: a)
$\mu \ne 0$ and $B=0$ (chiral symmetry breaking/restoration), b) $\mu=0$ and
$B \ne 0$ (magnetic catalysis) and c) $\mu \ne 0$ and $B\ne0$ (magnetic induced
quark matter stability). The quantity $E/A$, for magnetized quark matter
described by NJL SU(2) model,  has a minimum which is lower than the one
determined for magnetic free quark matter. 
We have also obtained that a magnetic field of the order
of $0.2 \times 10^{19}\,\,$G barely affects the effective mass as compared 
with the results for matter not subjected to the magnetic field and for 
$B=5\times 10^{19}$ G matter is totally polarized for chemical potentials 
below 490 MeV. For small values of the magnetic
fields the number of filled LL is large and the quantization effects are
washed out, while for large magnetic fields the chiral symmetry restoration
occurs for smaller values of the chemical potentials.

In order to introduce $\beta$-equilibrium, we
have extended the thermodynamical potential so as to consider leptonic
contributions. Our numerical results show that, for the SU(2) case, only very
high magnetic fields ($B \ge 10^{18} \,$G) affect the EoS in  a noticeable
way.

While for quark matter in $\beta$-equilibrium, the densities at which
the pressure is zero increase with the inclusion of the magnetic field,
the opposite happens in matter not subjected to $\beta$-equilibrium.

Regarding the TOV equation results, we can see that the SU(2) version with
magnetic field provides masses and radii for the
maximum mass stable configuration larger than the analogous magnetic free
configuration. The radii are quite lower than the corresponding
one obtained within the magnetic free SU(3) NJL. This is mainly due to
the softness of the SU(2) EoS at the lower densities.
The mass-radius results were obtained with  a density dependent magnetic field,
 which increases from the surface to the interior of the star.
 The masses of the maximum mass stable configuration for fields of the order
or larger than $10^{19}$ G are larger than 1.4 $M_\odot$, and much larger than 
the analogous magnetic free configuration,  due to the
contribution of the magnetic field.

The inclusion of the magnetic field in the SU(3) version of the NJL is not 
trivial and is currently under way.
We expect to obtain larger results for the radius.
Knowing that the effects of the anomalous magnetic moments is very 
relevant we also intend to take them into account in the next calculations.
An alternative model, usually called PNJL\cite{Scoccola} , can also be investigated in the presence of
magnetic fields.

\section*{Acknowledgments}

This work was partially supported by
 by CNPq (Brazil) by FCT (Portugal) under the projects
POCI/FP/81923/2007 and PTDC/FIS/64707/2006.
APM has been
supported by CITMA/Cuba under grant CB0407 and the ICTP
Office of External Activities through NET-35. APM acknowledges also
the support of TWAS-UNESCO and ICRA-CBPF Brazil and the hospitality
of Universidade Federal de Santa Catarina in Florianopolis.

\appendix

\section{Summing Matsubara frequencies and related formulas}

In this appendix we give the results for the main integrals and Matsubara sums
appearing along the text. The Matsubara sums which are relevant for the different integrals considered in our work can be derived as (see e.g. \cite{kapusta}):

$$ T \sum_{n=-\infty}^{+\infty}
\ln [(\omega_n-i \mu)^2 + E_p^2]$$
\begin{equation}
= E_p + T \ln\{1+ e^{-[E_p+\mu]/T}\}
+ T \ln\{1+ e^{-[E_p-\mu]/T}\}\;,
\label{sum1}
\end{equation}
which, as $T\to 0$, becomes
$$ \lim_{T \to 0} T \sum_{n=-\infty}^{+\infty}
\ln [(\omega_n-i \mu)^2 + E_p^2]$$
\begin{equation}
= E_p + [ \mu -E_p] \theta(\mu -E_p) = {\rm max}(E_p,\mu).
\label{sum1tzero}
\end{equation}

\section{Evaluation of Divergent Integrals}

Consider the divergent term
\begin{equation}
 \frac{N_c }{2\pi} \sum_{f=u}^{d} \sum_{k=0}^{\infty} (2-\delta_{k 0})(|q_f|B)\int_{-\infty}^{\infty} \frac{d p_z}{(2\pi)}E_{p,k}(B),
\label{start}
\end{equation}
appearing in Eq. (\ref {pressBmu}). By adding and subtracting a LLL term to it one gets
\begin{equation}
 \frac{N_c }{\pi} \sum_{f=u}^{d} \sum_{k=0}^{\infty} (|q_f|B)\int_{-\infty}^{\infty} \frac{d p_z}{(2\pi)}\left [E_{p,k}(B)- \frac{E_{p,0}(B)}{2} \right ].
\end{equation}
Now, the integrals can be performed in the following way. We change dimensions from $1 \to d=1-\epsilon$ use the
standard dimensional regularization formula \cite {ramond}
\begin{equation}
\int_{-\infty}^{\infty} \frac{d^d q}{(2\pi)^d}[q^2+M^2]^{-A}= \frac{\Gamma[A-d/2]}{(4\pi)^{d/2} \Gamma[A] (M^2)^{A-d/2}}
,
\end{equation}
obtaining, after defining $x=M^2/(2qB)$
\begin{eqnarray}
&&\frac{2N_c }{\pi} \sum_{f=u}^{d} \sum_{k=0}^{\infty} (|q_f|B)^2 \frac{\Gamma[-1+ \epsilon/2]}{(4\pi)^{1/2-\epsilon/2} \Gamma[-1/2]}\nonumber \\
&\times&
\left \{ \frac {1}{[k + x]^{-1+\epsilon/2}} - \frac{1}{2 x^{-1+\epsilon/2}} \right \}.
\end{eqnarray}
Then, using the definition of the Riemann-Hurwitz zeta function one can get rid of the summation over Landau levels obtaining
\begin{equation}
- \frac{N_c }{2\pi^2} \sum_{f=u}^{d}(|q_f|B)^2 \Gamma[-1+ \epsilon/2] \left [ \zeta(-1+\epsilon/2,x) - \frac{1}{x^{1-\epsilon/2}} \right ]\,\,.
\end{equation}
After expanding around $\epsilon=0$ and canceling few terms one has the (still divergent) contribution
\begin{equation}
- \frac{N_c }{2\pi^2} \sum_{f=u}^{d}(|q_f|B)^2 \left [ \frac{x^2}{\epsilon} + \frac{x^2}{2}(1-\gamma_E) - \frac{x}{2} \ln x - \zeta^\prime(-1,x) \right ].
\label{div1}
\end{equation}
One can now get rid of the above divergence by adding and subtracting  a vacuum term (see Eq. (FF))
\begin{equation}
P^{vac}= 2N_c N_f \int  \frac{d^3 {\bf p}}{(2\pi)^3}[{\bf p}^2 + M^2]^{1/2}.
\end{equation}
The subtracted term can be conveniently treated by performing a change of variables ${\bf p}^2 \to [{\bf p}^\prime]^2/(2qB)$ and $M^2 \to x=M^2/(2qB)$. Then replacing
$N_f$ by $\sum_{f=u}^d$ and performing the integration in $d=3-\epsilon$ dimensions leads to
\begin{equation}
- P^{vac}=\frac{N_c }{2\pi^2} \sum_{f=u}^{d}(|q_f|B)^2 \left [ \frac{x^2}{\epsilon} + \frac{x^2}{2}(1-\gamma_E) - \frac{x^2}{2} \ln x +\frac{x^2}{4} \right ]\,\,\,.
\label{div2}
\end{equation}
Finally, adding all contributions one may write Eq. (\ref {start}) as
\begin{equation}
P^{vac}+ \frac{N_c }{2\pi^2} \sum_{f=u}^{d}(|q_f|B)^2 \left [ \zeta^\prime(-1,x) - \frac{1}{2} (x^2-x) \ln x+ \frac{x^2}{4} \right ]\,\,\,,
\end{equation}
where the added quantity, $P^{vac}$, can be evaluated using a sharp cut off reproducing
the results quoted in the text, which are in agreement with Ref. \cite {klimenko03}. Also, note that our strategy avoids possible regularization complications introduced if one first performs a proper time type of calculation and then introduces a three momentum cut off.

 The following relation has also been used in the above derivation \cite {wolfram}
\begin{equation}
\zeta(-1,x) = - \frac {B_2(x)}{2} = -\frac{1}{2}\left [ \frac{1}{6} - x + x^2 \right ],
\end{equation}

where $B_n(x)$ represents the Bernoulli polynomial.
Further, to obtain the gap equation for the $B \ne 0$ case the following relations \cite {wolfram} are useful, $d \zeta[0,x]/dz = \ln \Gamma(x) - (1/2) \ln (2 \pi)$, $\zeta[0,x]=1/2 -x$, and $d \zeta[z,x]/dx= -z \zeta [z+1,x]$.

\end{document}